\documentclass[%
twocolumn,
superscriptaddress,
amsmath,amssymb,
aps,
]{revtex4-2}

\usepackage{graphicx}
\usepackage{bm}
\usepackage{hyperref}
\usepackage{color} 
\usepackage{txfonts}
\usepackage[dvipsnames]{xcolor}
\usepackage{ragged2e}
\usepackage{amsmath}

\begin{document}

\title{Quantum control of classical motion: piston dynamics in a Rabi-coupled Bose-Einstein condensate}

\author{Jing Li}
 \affiliation{School of Physics, University College Cork, T12 H6T1 Cork, Ireland}

\author{E. Ya. Sherman}%
\affiliation{Department of Physical Chemistry, University of the Basque Country UPV/EHU, 48940 Leioa, Spain}%
\affiliation{IKERBASQUE, Basque Foundation for Science, 48009 Bilbao, Spain}
\affiliation{EHU Quantum Center, University of the Basque Country UPV/EHU, 48940 Leioa, Spain}

\author{Andreas Ruschhaupt}
\affiliation{School of Physics, University College Cork, T12 H6T1 Cork, Ireland}%

\begin{abstract}
We explore the dynamics of a hybrid classical-quantum system consisting of a classical piston and a 
self-interacting pseudospin 1/2 Bose-Einstein condensate with a time-dependent Rabi coupling. 
We investigate the mechanical work produced by the piston moving as a result of the quantum pressure
of the condensate. The time-dependent Rabi field redistributes the condensate density between the spin components and, as a result, causes a time-dependent pressure acting on the piston. Correspondingly, the motion of the piston produces quantum evolution of the condensate mass- and spin density profiles.
We show how by optimised design of the time-dependent direction of the Rabi field, one can control position and velocity of the piston. 
\end{abstract}

\maketitle

\section{Introduction}
\label{sec:1}

Classical and quantum heat machines are important for both fundamental research and applications in technologies \cite{Blickle2012,Goold_2016,Stevebook}. 
The analysis of the operation of classical heat machine
\cite{callen_textbook}
is based on applications of equation of state of the classical working medium relating its pressure $P$, 
volume $V$, and temperature $T.$  This is usually done in terms of equations of state
with the most well-known example for the ideal gas in the form $PV=\nu T,$
where $\nu$ is the number of the mole units.

In classical heat engines, another important object is the "piston" which is connected to the working medium.
The temperature- and volume-related pressure determines 
the motion of the piston attached to the classical matter, 
producing a mechanical work dependent on the time-dependent position and velocity of the piston.
This mechanical work of the piston is essential in the study of classical heat engines
as, for example, in determining their efficiencies.

Recently, also quantum heat engines have been demonstrated across various quantum platforms, 
including trapped ions \cite{Lutz2012,Kilian2016},
ultracold atoms \cite{Widera21},
quantum dots \cite{Sothmann_2014,Franco2021} and
optomechanical oscillators \cite{Pierre2014,Markus2014,Elouard_2015,Gaub2002}.
Another option is to consider atomic Bose-Einstein condensates including large number of atoms 
as the working medium \cite{Charalambous_2019,Myers_2022,Li2018,Niedenzu2019}. 
Condensates demonstrate properties related to interatomic interactions
of various symmetries and spin-related physics with Rabi (Zeeman-like) and spin-orbit coupling interactions which can be taken advantage of in quantum heat engines, see for example \cite{Jing22}. In addition, a quantum many-body engine has been shown in \cite{Koch2022} in the BEC-BCS crossover by changing a magnetic field. 
Various methods of quantum optimal control have been also applied in quantum heat engines, such as for example shortcuts to adiabaticity technique \cite{TORRONTEGUI2013117,STAreview}
and optimal control schemes \cite{Glaser2005,Montangero2011,Glaser2015,Kiely2023}.

However, one faces the issues that definition of the work produced by 
the quantum system might be non trivial \cite{KieuPRL04,Hanggi07,TalknerRMP2011,Molmer2020}.
Moreover, in some of these systems one cannot have a well-defined size and, 
therefore, cannot use a piston-like description, which is an essential element of a thermal machine.
Therefore, the piston in quantum heat engines might need to be studied in detail \cite{Kurizki14}.

All these have motivated this paper in which we study the "piston" in the quantum setting
of a Bose-Einstein condensate as a working medium with a time-dependent Rabi field. The BEC is coupled to a "classical"
piston. We keep the physical implementation of the "classical" piston in general. However, for example,
the piston can be potentially produced by a particle held by an optical tweezer \cite{Ashkin1971,Ashkin1986}.
In our system, the Rabi field redistributes the condensate density between the spin components and,
as a result, causes a time-dependent pressure acting on the piston, while the motion 
of the piston produces also a quantum evolution of the condensate.
The first goal of this paper is that we are interested in a clear definition of the work done by the piston
in this setting; this will be also useful in quantum heat engines in general.
Furthermore, the second goal, even more general, is that we examine and optimise how the trajectory of the "classical" piston can be controlled
via a control of the quantum system. In detail, we consider the control by acting on the spin degree of freedom
by the time-dependent  direction of the Rabi field. Changing this field causes redistribution of the spin self-interacting components
and, thus, modifies the pressure and moves the piston, producing a \textit{classical} 
mechanical work. Here we use a two-step optimization scheme for quantum control of the BEC which will result
in the required piston motion in terms of position and velocity achieved at a certain evolution time.

This paper is organized as follows. In Sec. \ref{sec:2} we formulated the model describing the BEC
and the piston and show how to consider classical motion of the piston as the 
limit of its quantum motion. In Sec. \ref{sec:3} we study properties of the stationary BEC and show
how the pressure depends on the Rabi field. This dependence allows one to control the 
piston motion. Furthermore, we study a two-step optimization scheme for quantum control of the piston motion
in terms of position and velocity achieved at a certain evolution time. In Sec. \ref{sec:4} we discuss 
possible relation to experiment and formulate the conclusions of this paper.

\section{Model: BEC and Piston}
\label{sec:2}

\subsection{Equations of motion}
\vspace{-0.1cm}

\begin{figure}[t]
\begin{center}
\includegraphics[width=0.80\columnwidth]{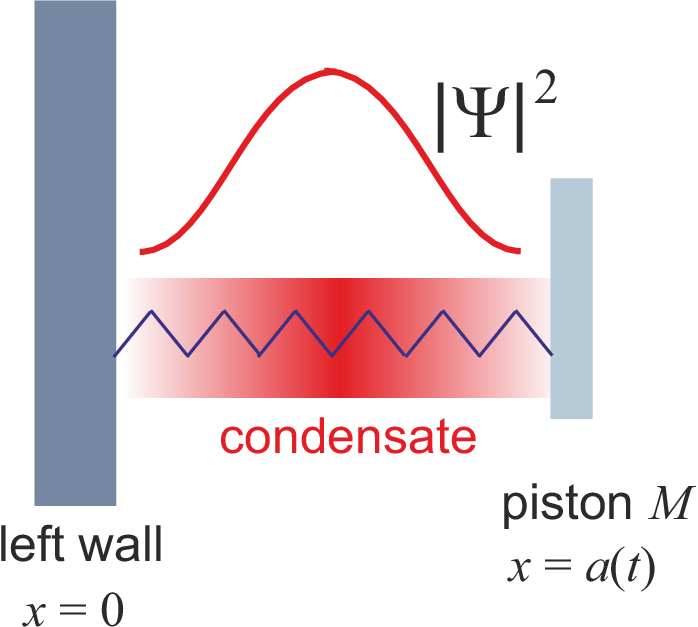}
\end{center}
\vspace*{-5mm}
\caption{Systematic scheme: A two-component BEC is confined between a stationary wall (left) at $x=0$ and a moving piston (right) at $a(t)$. 
The piston is interacting with BEC as well as an additional external harmonic potential (spring).
}
\label{fig1:piston}
\end{figure}

We consider a two-component BEC confined by an external potential and a piston, see Fig. \ref{fig1:piston}.
The piston is interacting with the BEC as well as an additional external harmonic potential.
As already mentioned, we will keep the physical implementation of the piston general; however, we have in mind for example
the piston can be possibly produced by a particle hold by an optical tweezer \cite{Ashkin1971,Ashkin1986}.  
We also consider a one-dimensional spatial approximation, i.e. we are considering only the longitudinal $x$ direction and we assume a strong vertical confinement in $y$ and $z$ directions.
The time $t-$ and position $x$-dependent wavefunction of BEC has the form ${\bm \Psi}(x,t) \equiv (\psi_{\uparrow}(x,t),\; \psi_{\downarrow}(x,t))^{\rm T}$ (here ${\rm T}$ stands for transposition) and $\psi_{\uparrow}(x,t)$ and $\psi_{\downarrow}(x,t)$ are related to the two pseudo-spin components. In what follows, for brevity, we omit when possible the explicit $(x,t)-$dependence. The total Hamiltonian is given by
\begin{eqnarray}
   \hat H = \hat H_{B} + \hat H_{p} +\hat H_{Bp},
   \label{hamfull}
\end{eqnarray}
with $\hat H_{B}$ describing the BEC system, $\hat H_{p}$ describing the quantum piston and $\hat H_{Bp}$ describing the coupling between BEC and piston. In detail, the Hamiltonian $\hat H_{B}$ of the spin-1/2 BEC is given by
\begin{eqnarray}
\hat H_{B} =\frac{\hat{p}_x^{2}}{2m} +\frac{\hbar}{2} 
    {\bm \Delta}\cdot\hat{\bm\sigma}
+ \hat G + V_{L}(x), 
\label{ham0}
\end{eqnarray}
with $\hat{p}_{x}=-i\hbar\partial_{x}$ being the momentum operator in the longitudinal $x$ direction,  particle mass $m$, the Rabi field ${\bm \Delta}=\Delta(0,\sin\phi,\cos\phi),$ ${\bm \sigma}=(0,\hat{\sigma}_{y},\hat{\sigma}_{z}),$ and $\hat{\sigma}_{y,z}$ being the Pauli matrices.
The operator $\hat G$ is $2\times2$ matrix with
components $G_{11}=g_{s}\vert\psi_{\uparrow}\vert^{2}+g_{c}\vert\psi_{\downarrow}\vert^{2} $ and $G_{22}=g_{s}\vert\psi_{\downarrow}\vert^{2}+g_{c}\vert\psi_{\uparrow}\vert^{2}$ and $G_{12}=G_{21}=0.$ 
The parameters
$g_{s}=\pi\hbar^{2} a_{s}/{m}$ and $g_{c}=\pi\hbar^{2} a_{c}/{m}$ represent the strength of the self-interaction and cross-interaction. The $s-$wave scattering lengths $a_{s}$ and $a_{c}$ can be tuned by Feshbach resonance \cite{Feshbach98}. Note that $g_{s} =g_{c}$ corresponds to Manakov’s symmetric system \cite{Manakov1974}. The parameters $\Delta$ and $\phi$ are the strength and the angle of the Rabi-like magnetic field. 
The condensate is confined on one side by a very high step-like narrow left barrier $V_{L}(x)$
sufficient to suppress the condensate density at $x<0.$
The piston is connected to a spring such that the Hamiltonian of the piston is
\begin{equation}
\hat{H}_p = \frac{\hat{p}_{X}^{2}}{2M} + \frac{1}{2} M \Omega^{2} (\hat{x}-X_{0})^{2},
\end{equation}
where $M$ being the mass of piston, and $\hat{p}_X=-i\hbar\partial_{X}$ being the corresponding momentum operator. Such a potential for the piston can be produced by using optical tweezers capable of holding particles of different masses and sizes \cite{Ashkin1986}.
The frequency of harmonic oscillator is $\Omega =\sqrt{k/M}$, where $k$ is the spring constant, and $X_{0}$ is the equilibrium position of the spring. 
The interaction between the BEC and the piston is given by
\begin{equation}
    \hat{H}_{Bp} = \hat{V}(\hat{x}-\hat{X}).
    \label{eq:HBp}
\end{equation}

Applying Ehrenfest's theorem for the system, we have the instantaneous change in the expectation value of the position $\hat{x}$:
\begin{equation}
\label{eq:dXdt}
\frac{d}{dt} \langle \hat{x} \rangle = \frac{1}{i\hbar}\langle [\hat{x}, \hat H] \rangle= \frac{1}{i\hbar}\langle [\hat{x}, \hat H_{p}] \rangle=\frac{\langle \hat{p}_X \rangle}{M}.
\end{equation}
Similarly, we have 
\begin{equation}
\label{eq:dpdt}
    \frac{d}{dt} \langle \hat p_X \rangle
    = -M \Omega^{2}(X_{c}(t)-X_{0}) +\frac{1}{i\hbar}\langle [\hat p_X, \hat V] \rangle.
\end{equation}
We now do the variable change $\hat\zeta = \hat{x} - a(t)$ where $a(t) = \langle \hat{X}\rangle.$  It follows that $\langle \hat{\zeta}\rangle=0.$

Assuming that $\zeta$ is small, $\hat{H}_{Bp}$ in Eq. \eqref{eq:HBp} can be approximately written as
\begin{equation}
\label{eq:operatorV}
    H_{Bp}=\hat V (\hat{z}) \approx V(\hat{x} - a(t)) - \hat\zeta \frac{d V}{dz}(\hat{x} - a(t)),
 + {\mathcal O} (\zeta^{2})   
\end{equation}
where $\hat{z} = \hat{x} - a(t) - \hat{\zeta}.$ 
By substituting Eq. (\ref{eq:operatorV}) into Eq. (\ref{eq:dpdt}), we obtain
\begin{equation}
\label{eq:dpdtnew}
    \frac{d}{dt} \langle \hat p_X \rangle = -M \Omega^{2}(a(t)-X_{0}) + P(t) + {\mathcal O} (\zeta^{2}),
\end{equation}
where the exact "pressure" of the BEC  (which in a one-dimensional system is analogous to a generalized force) is determined by the instantaneous wavefunction ${\bm\Psi}$ in the vicinity of the piston and given by:
\begin{eqnarray}
   P(t) &=& \Big\langle {\bm \Psi}\Big| \frac{dV}{dz}\left(\hat{x} - a(t)\right) \bm\Psi \Big\rangle \nonumber\\
   &=&\int^{+\infty}_{-\infty} dx \; \vert \bm\Psi\vert^{2} \frac{d V(x-a(t))}{d x}.
   \label{eq:pressure}
\end{eqnarray}
where ${\bm \Psi}$ is a solution of the  Gross-Pitaevskii equation of 
the BEC Hamiltonian $\hat H_{B}$ alone! 
The coupled Gross-Pitaevskii equations of $\hat H_{B}$ alone are then given by
\begin{eqnarray}
i\hbar\frac{\partial }{\partial t}\psi_{\uparrow} &&=\left( -\frac{\hbar^{2}}{2m}\frac{\partial ^{2}}{%
\partial x^{2}}+\frac{\hbar}{2}\Delta \cos\phi(t) + g_{s} \vert\psi_{\uparrow}\vert^{2} + g_{c}\vert\psi_{\downarrow}\vert^{2}\right) \psi_{\uparrow} %
 \nonumber \\
&&+\left[V_{L}(x)+V(x-a(t))\right]\psi_{\uparrow}
- i \frac{\hbar}{2}\Delta \sin\phi(t)\;\psi_{\downarrow},
  \nonumber\\
i\hbar\frac{\partial}{\partial t}\psi_{\downarrow} &&=\left( -\frac{\hbar^{2}}{2m}\frac{\partial ^{2}}{%
\partial x^{2}} - \frac{\hbar}{2} \Delta \cos\phi(t)
+g_{s}\vert\psi_{\downarrow}\vert^{2}+g_{c}\vert\psi_{\uparrow}\vert^{2}\right) \psi_{\downarrow}
   \nonumber \\
&&+\left[V_{L}(x)+V(x-a)\right]\psi_{\downarrow} + i \frac{\hbar}{2} \Delta \sin\phi(t)\;\psi_{\uparrow},
\label{coupled_GPE}
\end{eqnarray}
and $\int_{-\infty}^\infty dx\, \vert \bm\Psi\vert^{2} = \int_{-\infty}^\infty dx\, \left(\vert\psi_{\uparrow}\vert^{2} + \vert\psi_{\downarrow}\vert^{2}\right) = N$ with $N\gg\,1$ being the particle number of the BEC.
In the following, we neglect terms ${\mathcal O} (\zeta^{2})$ and
the equations of motion for a quantum piston shown in Eqs. (\ref{eq:dXdt}) and (\ref{eq:dpdtnew}) can then be combined to give
\begin{eqnarray}
M \frac{d^{2}}{dt^{2}} a(t) = -M \Omega^{2}a(t) + P(t),
\label{eq:Newton}
\end{eqnarray}
where we have also set $X_{0}=0.$
This shows that the piston position $a(t)$ depends both on the spring and on the BEC system with $P(t)$ being given by Eq. \eqref{eq:pressure}.

\subsection{Mechanical Piston Work}

Let us now consider the mechanical work done by the piston. In general, this work can be calculated as
\begin{eqnarray}
    W_{p} = \int_{0}^{t_f} dt\; \frac{da}{dt} P(t)\,.
    \label{eq:work}
\end{eqnarray}

As the first step, we define by $P_{\rm st} (a,{\bm\Delta})$ the stationary pressure as
\begin{equation}
P_{\rm st} (a,{\bm\Delta}) = -\frac{\partial E_{\rm gs} (a,{\bm\Delta})}{\partial a},
\label{PdE}
\end{equation}
where $E_{\rm gs}(a,{\bm\Delta})$ is the ground state energy 
of the BEC (related to stationary version of Eqs. \eqref{coupled_GPE})
with given piston position $a$, angle $\phi$ and detuning $\Delta.$

We can now rewrite the exact ${\bm\Psi}-$determined 
pressure as $P(t) \equiv P_{\rm st} (a(t),{\bm\Delta}) + \delta P(t),$ 
where $\delta P(t)$ is a relatively small (as supported by our following 
calculations) non-stationary contribution.
Then, we obtain
\begin{equation}
    W_p = W_{p,{\rm st}} + \delta W_{p}
\end{equation}
where the stationary work contribution is
\begin{eqnarray} \label{eq:workapprox}
    W_{p,{\rm st}} &=& \int_{0}^{t_f} dt\; \frac{da}{dt} P_{\rm st}(a(t),{\bm\Delta}(t)) \\
   & = & - \int_{0}^{t_f} dt\; \frac{da}{dt} 
   \frac{\partial}{\partial a} E_{\rm gs} (a(t),{\bm\Delta}(t)).\nonumber
\end{eqnarray}
and the non-stationary work contribution is
\begin{eqnarray}
\label{dWp}
 \delta W_{p} = \int_{0}^{t_f} dt\; \frac{da}{dt} \delta P(t).  
\end{eqnarray}

From now on, if not stated otherwise, we assume that the time dependence ${\bm\Delta}=\Delta(0,\sin\phi(t),\cos\phi(t))$
is due solely to the time-dependent angle $\phi(t)$. Thus, the full time-derivative of 
$(a(t),\phi(t))-$ dependent energy $E_{\rm gs}$ can be written as
\begin{eqnarray}
  \frac{dE_{\rm gs}}{dt}= \frac{\partial E_{\rm gs}}{\partial a} \frac{da}{dt} 
  + \frac{\partial E_{\rm gs}}{\partial \phi} \frac{d\phi}{dt}\,.
\label{eq:dEdt}
\end{eqnarray}
By using Eq. \eqref{eq:dEdt} we get for the stationary term,
\begin{equation}
 W_{p,{\rm st}}=- {\mathcal E}_{\rm st} + W_{\phi,{\rm st}}\, ,
\end{equation}
where the total stationary energy change is
\begin{eqnarray}
{\mathcal E}_{\rm st} &\equiv& E_{\rm gs} (a(t_f),\phi(t_f)) - E_{\rm gs}(a(0),\phi(0)) \\
&=& \int_{0}^{t_f} dt\; \frac{d}{dt} E_{\rm gs} (a(t),\phi(t)). \nonumber
\end{eqnarray}
The work related to the change of $\phi$ becomes:
\begin{equation}
     W_{\phi,{\rm st}} = \int_{0}^{t_f} dt\; \frac{d\phi}{dt} \frac{\partial}{\partial \phi} 
     E_{\rm gs} (a(t),\phi(t))\, .
\end{equation}
In summary, we have shown that the mechanical work done by the piston can be split up in the following components:
\begin{eqnarray}
     W_{p} = -{\mathcal E}_{\rm st} + W_{\phi,{\rm st}} + \delta W_{p}\, ,
\end{eqnarray}
which can be also seen that the work done on the system by changing $\phi(t)$ is split
into three components:
\begin{eqnarray}
  W_{\phi,{\rm st}} = W_{p} + {\mathcal E}_{\rm st} - \delta W_{p}\, ,
  \label{workcomponents}
\end{eqnarray}
where $\delta W_{p}$ is the non-stationary contribution as determined by Eq. \eqref{dWp}. 
We will examine these different contributions in the example in  Sect. \ref{sect3:opt}.

\subsection{Interplay of the Rabi field, self-interaction, and the pressure}
\label{sec:trial}

We first consider a simple mean-field model which permits one to understand qualitatively the role of the Rabi coupling in the self-interacting BEC pressure. In this subsection, we assume that the piston potential is infinite.
We consider an approximate trial wavefunction of the system, Eq. (\ref{hamfull}) with $\phi=0$ and $\Delta>0$ as
\begin{equation} 
    {\bm\Psi}_{\rm tr}(x)=\sqrt{\frac{2}{a}}\sin(\pi\,x/a)
\begin{bmatrix}
\sqrt{N_{\uparrow}} \\
\sqrt{N_{\downarrow}} 
\end{bmatrix},
\end{equation}
where $N_{\uparrow}+N_{\downarrow}=N.$   
The nonlinear self- and cross- interaction energy is given by 
\begin{equation}
E_{\rm sc}=\int^{+\infty}_{-\infty}dx\,
\left[
\frac{g_{s}}{2}\left(|\psi_{\uparrow}|^{4}+ |\psi_{\downarrow}|^{4}\right)+
g_{c}|\psi_{\uparrow}|^{2}|\psi_{\downarrow}|^{2}
\right];
\label{nonE}
\end{equation}
it depends only on the total number of particles at $g_{c}=g_{s}.$
Below we consider the case with $g_{c}=0.$
The spin-related Rabi coupling-related energy reads
\begin{equation}
E_{R} = \frac{\Delta}{2}(N_{\uparrow}-N_{\downarrow})=\frac{\Delta}{2}NS,
    \label{ZeemanE}
\end{equation}
where the stationary magnetization $S_{\rm st}$ is defined by
\begin{equation}
    S_{\rm st}\equiv\frac{1}{N}\left(N_{\uparrow}-N_{\downarrow}\right).
\end{equation}
By using Eq. \eqref{nonE} the total energy $E_{\rm tr}(a,\Delta)$ 
in terms of $N_{\downarrow}$ can be written as
\begin{equation}
    E_{\rm tr}(a,\Delta) = \frac{\pi^{2}}{2a^{2}}\frac{\hbar^{2}}{m}N+\frac{3g_{s}}{4a}N^{2}+\frac{\Delta}{2}N
    -\left(\frac{3g_{s}}{2a}N+\Delta\right)N_{\downarrow}    
    +\frac{3g_{s}}{2a}N_{\downarrow}^{2}.
\label{totE1}
\end{equation}
Thus, the Rabi coupling tends to accumulate particles 
in the spin-down states while self-repulsion favours equal distribution between the spin states. Interplay of these two effects determines the distribution of particles and the spin polarization and in the presence of a non-uniform external potential they lead to a motion of soliton \cite{Mardonov2019,Eichmann2021,Sarkar2023}.
The equilibrium condition yields stationary $N_{\downarrow}=N/2+a\Delta/(3g_{s})$ and
shows that the state is fully spin-down polarized when $N_{\downarrow}=N$ with
\begin{equation}
  a\Delta=\frac{3}{2}Ng_{s}.
  \label{critad}
\end{equation}
When $\Delta=0$, $N_{\downarrow}=N_{\uparrow} = 1/2$, and the total ground state 
energy from Eq. (\ref{totE1}) is
\begin{equation}
  E_{\rm tr,gs}(a,0)=\frac{\pi^{2}}{2a^{2}}\frac{\hbar^{2}}{m}N
  +\frac{3g_{s}}{8a}N^{2}. 
\label{eq:EDelta0}  
\end{equation}
When $\Delta>\Delta_{\rm cr}$ with $\Delta_{\rm cr}=3Ng_{s}/(2a)$ 
(or, correspondingly, $a_{\rm cr}=3Ng_{s}/(2\Delta)$), the BEC is fully spin-polarized with the total energy
\begin{equation}
  E_{\rm tr,gs}(a,\Delta>\Delta_{\rm cr})=E_{\rm tr,gs}(a,0)+\frac{3g_{s}}{8a}N^{2}-\frac{\Delta}{2}N,
\end{equation}
and in the intermediate region $\Delta<\Delta_{\rm cr}$ we obtain
\begin{equation}
\label{Delta_interm}
    E_{\rm tr,gs}(a,\Delta<\Delta_{\rm cr})=E_{\rm tr,gs}(a,0)-\frac{1}{6}\frac{a\Delta^{2}}{g_{s}},
\end{equation}
corresponding to $a-$dependent $\sim\,\Delta^{2}$ energy decrease due to the Rabi coupling.

The stationary pressure $P_{\rm tr,st}(a,\Delta)$ has the 
form defined by Eq. \eqref{PdE} corresponding to the energy $E_{\rm tr,gs}(a,\Delta).$
The pressure difference in the fully polarized BEC due to the Rabi coupling is
\begin{equation}
    p_{\rm tr,st}(a,\Delta) \equiv P_{\rm tr,st}(a,\Delta)-P_{\rm tr,st}(a,0)=\frac{3g_{s}}{4a^{2}}, \quad \Delta>\Delta_{\rm cr}, 
    \label{Eq:dp1}
\end{equation}
while at $\Delta<\Delta_{\rm cr},$ the pressure shift is
\begin{equation}
    p_{\rm tr,st}(a,\Delta) \equiv P_{\rm tr,st}(a,\Delta)-P_{\rm tr,st}(a,0)=\frac{\Delta^{2}}{6g_{s}}, \quad \Delta<\Delta_{\rm cr}.
    \label{Eq:dp2}
\end{equation}
For this trial function, the equilibrium position of the piston $a_{\rm eq}$ is determined by condition $P_{\rm tr,st}(a,\Delta)=k a_{\rm eq}.$ Thus, the $\Delta-$dependence of the pressure 
makes Rabi-coupled BECs suitable for producing mechanical
work by using a time-dependent Rabi coupling.

\begin{figure}[t]
{\includegraphics[width=0.85\columnwidth]{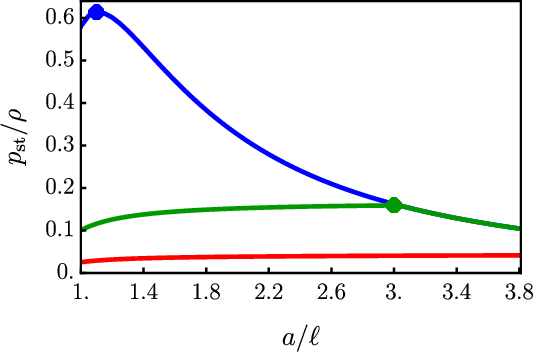}}
\caption{The $\Delta$-dependent pressure shift versus piston
position $a$ for $\Delta\tau=5$ (blue), $\Delta\tau=2$ (green) and $\Delta\tau=1$ (red) in Eqs. (\ref{Eq:dp1}) and (\ref{Eq:dp2}) with fixed $ \phi=0$. Other parameters here and in the following figures: $g_{s}/(\epsilon\ell)=5, g_{c}=0, V_{0}/\epsilon=10, V_{L}/\epsilon=100, s/\ell=0.1.$   
 }
\label{fig:test}
\end{figure}

\begin{figure}
{\includegraphics[width=0.85\columnwidth]{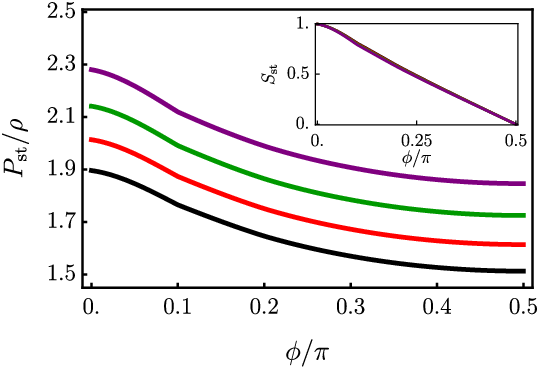}}
\caption{Stationary pressure $P_{\rm st}(a,\phi)$ with respect to the angle $\phi$ and for different piston positions $a/\ell=1.65$ (purple), $a/\ell=1.7$ (green), $a/\ell=1.75$ (red), and $a/\ell=1.8$ (black). Inset: The stationary magnetisation $S_{\rm st}$ defined for the corresponding wavefunction. $\Delta\tau = 5$. }
\label{fig:P_phi}
\end{figure}


\begin{figure}[h]
{\includegraphics[width=0.85\columnwidth]{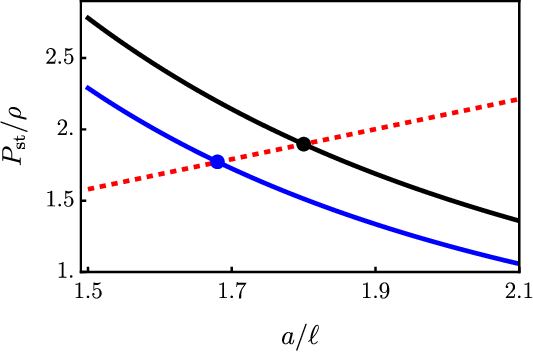}}
\caption{Stationary pressure $P_{\rm st}(a,\phi)$ with respect to the piston position $a$ with $\phi=0$ (solid black) and $\phi=\pi/2$ (solid blue).  
Straight dotted 
red line corresponds to the force $ka$ with $k\ell/\rho=1.05$.
}
\label{fig:fixed piston}
\end{figure}

\section{Quantum control of the piston motion}
\label{sec:3}

 We begin by considering a static piston which will serve as a reference point
 and then a time-dependent situation where the piston can move under the influence of the BEC- and laser-produced forces. The goal is to control the piston motion by acting on the quantum degree of freedom 
of the BEC by changing in time the angle $\phi(t).$ 

In this section, we consider more realistic smooth potential functions as follows:  
\begin{eqnarray}
\label{eq:newpot}
V_{L}(x)&=& V_{L}\times \left\{
\begin{array}{l}
1-\theta\left(x-s\right), \hspace{1.9cm} {|x| > s}\\
\\
\left[1-\sin({\pi x}/(2s))\right]/2, \hspace{0.8cm} {|x| \leq s}
\end{array} \right.
\end{eqnarray}
for the left wall and 
\begin{eqnarray}
\label{eq:newpot_pist}
V(x)&=&V_{0}\times\left\{
\begin{array}{l}
\theta\left(x-s\right),\hspace{2.4cm} {|x| > s}\\
\\
\left[1+\sin({\pi x}/(2s))\right]/2, \hspace{0.8cm} { |x| \leq s}
\end{array} \right.
\end{eqnarray}
for the piston, where $\theta\left(x\right)$ is the Heaviside function, 
$V_{L}\gg V_{0}$ and the slope parameter $s$ being small compared to the piston position $a$.

Let us introduce also an experiment-related length unit $\ell$ and the corresponding related
energy unit $\epsilon\equiv \hbar^{2}/(m \ell^{2})$, the time unit $\tau \equiv m \ell^{2}/\hbar$ and the pressure (force) unit $\rho \equiv N\hbar^{2}/(m \ell^3)$.
In the following, we present our results in terms of these units in a dimensionless way to make our results general and,
as an example, we choose the parameters 
$M/(Nm)=1000$, $V_{L}/\epsilon=100$, $V_{0}/\epsilon=10$, $s/\ell=0.01$, $g_{s} N/(\epsilon\ell)=5$ 
and $g_{c}=0$ for the following analysis.
We will discuss some potential mappings to physical units in the conclusion section.

\subsection{Static piston and equilibrium states}

We begin with the static piston corresponding to its infinite mass and obtain the ground state
of the condensate ${\bm\Psi}_{\rm st}(x)$ corresponding to Eqs. \eqref{coupled_GPE},
\eqref{eq:newpot} and \eqref{eq:newpot_pist} 
by  the imaginary time evolution approach and then use Eq. \eqref{eq:pressure} to calculate the 
corresponding stationary pressure $P_{\rm st}$ for given system parameters as will be specified 
for the cases of interest. 
The first example we consider corresponds to the $a-$dependence of the pressure
at $\phi=0,$ making the energy and the pressure solely $\Delta-$dependent, 
for different values of $\Delta.$ Figure \ref{fig:test} shows the pressure difference 
$p_{\rm st}(a,\Delta)\equiv P_{\rm st}(a,\Delta)-P_{\rm st}(a,0)$ as a function of 
the length $a$ (cf. Eqs. \eqref{Eq:dp1} and \eqref{Eq:dp2}).
We can see that it is approximately a constant and then decreases with $a$. The critical length $a_{\rm cr}$ corresponds to the merging points which are denoted by the blue and green dots in Fig. \ref{fig:test},
where the effect of the Rabi field matches the effect of self-interaction, and a fully spin-polarized state
emerges, in agreement with the analysis in subsection \ref{sec:trial}. 

Next, in Fig. \ref{fig:P_phi}, we present the stationary pressure $P_{\rm st}(a,\phi)$ with respect  to the angle $\phi$ and for different fixed values of piston position
$a$.
It shows that $\phi=0$ corresponds to the maximum pressure with fixed length $a.$ 
At  $\phi=\pi/2$ both spin components are equally occupied, the self-interaction decreases, and the pressure becomes $\Delta-$independent. To add to this dependence, 
the stationary pressures $P_{\rm st}(a,\phi)$ for $\phi=0$ and $\phi=\pi/2$ 
are shown in Fig. \ref{fig:fixed piston} identifying the difference in the piston equilibrium position
in the presence of the external parabolic potential. 

\subsection{Control Goal}

The piston is connected to the spring which exerts a force $ka$ (dotted red line), where two dots present the equilibrium positions of piston in Fig. \ref{fig:fixed piston}.
When $\phi=0,$ the equilibrium length of the condensate is  $a_{\rm eq}(0)/\ell=1.80$.
We choose this equilibrium length as the starting point of the time evolution.
The equilibrium length for $\phi=\pi/2$ is $a_{\rm eq}(\pi/2)/\ell=1.68.$ The control target is to bring the 
piston to this equilibrium position at the final time $t_f$ with $\phi(t_f)=\pi/2.$ 
We also require that the velocity of the piston $v(t_{f})\equiv a'(t_{f})$
be zero at time $t_f$, i.e. $a'(t_f) = 0.$ So the piston should
end at the same position and with the same velocity that is achieved in a very slow adiabatic process, however in a finite time $t_f$.
Note that we are only interested in the piston, the BEC might not be
in a ground state at the end.

Corresponding to the control goal mentioned above, we are considering the following error function
\begin{eqnarray}
\xi^{2} = \left(\frac{a(t_{f})-a_{\rm eq}(\pi/2)}{a_{0}}\right)^{2}+\left(\frac{a'(t_{f})}{v_{0}}\right)^{2}
\label{xi}
\end{eqnarray}
where we choose $a_{0}/\ell=0.1$ and $v_{0} \tau/\ell=0.01$. We are aiming for an error $\xi^2 < 10^{-4}$ here.

\subsection{Reference control function}
As a reference case for the control of the piston motion, 
we assume a time-dependent control function $\phi(t)$ 
given by a third order polynomial in $t$ determined by fulfilling the following boundary conditions: 
\begin{equation}
\label{boundary1}
\phi(0)=0, \quad \phi(t_{f})=\pi/2,\quad \phi'(0)=\phi'(t_{f})=0,  
\end{equation}
so that the time-dependent $\phi(t)$ is given by
\begin{equation}
\phi_{\rm ref}(t)=\frac{3\pi}{2} \left(\frac{t}{t_{f}}\right)^{2}-\pi \left(\frac{t}{t_{f}}\right)^3.
\end{equation}
This reference control function for the example of $t_f/\tau=60$ is shown in Fig. \ref{fig5:examples}(a) (dashed, blue line).
The corresponding dynamics of the piston is obtained by numerically solving Eqs. \eqref{coupled_GPE} and Eq. \eqref{eq:Newton}.
The result for the same example is shown as the blue-solid line in Fig. \ref{fig5:examples}(b). It can be already seen from this figure that
the goal of $a(t_f)/\ell=1.68$ and $v(t_f)=0$ is not achieved by the reference control function, in detail, the error $\xi^2 = 0.69$. The corresponding magnetisation $S$ for this case can be seen in Fig. \ref{fig5:examples}(c) (solid, blue line).

It is interesting to compare the exact dynamics of the piston obtained by solving Eqs. (\ref{coupled_GPE}) and Eq. (\ref{eq:Newton}) numerically
with using an approximation. In detail, we approximate the pressure in the equation of motion 
for the piston by the stationary value $P_{\rm st}(a(t),\phi(t))$, i.e. we get
\begin{eqnarray}
M \frac{d^{2}}{dt^{2}} a(t) = -M \Omega^{2}a(t) + P_{\rm st}(a(t),\phi(t)),
\label{approx}    
\end{eqnarray}
and then we need only to solve numerically
the equation of motion of the piston function. This is much faster than solving Eqs. (\ref{coupled_GPE}) and Eq. (\ref{eq:Newton}) simultaneously.
This approximation also corresponds to assuming that the BEC is at any time $t$ in the ground state with corresponding piston position $a(t)$.
For the example of $t_f/\tau=60$, the resulting approximated piston trajectory is shown in \ref{fig5:examples}(b) (dashed-blue line).
This trajectory turns out as a good approximation of the exact piston trajectory (solid-blue line).
The magnetization $S$ corresponding to the two cases is shown in Fig. \ref{fig5:examples}(c) and the two blue lines are on top of each other.
Thus we find that this approximation works well and will be important in the next subsection.

\subsection{Optimisation of the control\label{sect3:opt}}

The goal here is to achieve the equilibrium length at the final time as well as at zero piston velocity.
Based on the finding of the last subsection, we will optimise the control function in the same way that we optimised the function for the approximated case and then apply it to the exact case.

\begin{figure}[t]
\hspace*{0.1cm}\includegraphics[width=0.85\columnwidth]{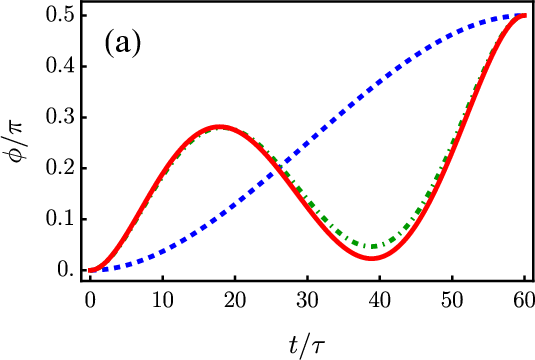}\\
\hspace*{0.0cm}\includegraphics[width=0.86\columnwidth]{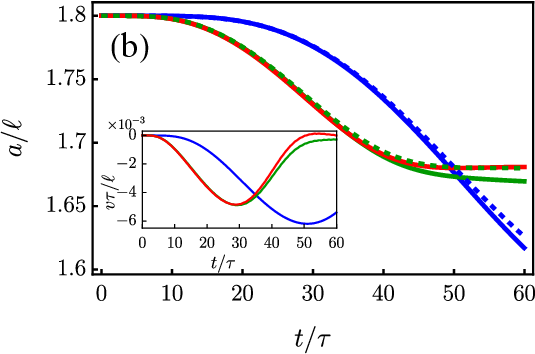}\\
\hspace*{0.1cm}\includegraphics[width=0.84\columnwidth]{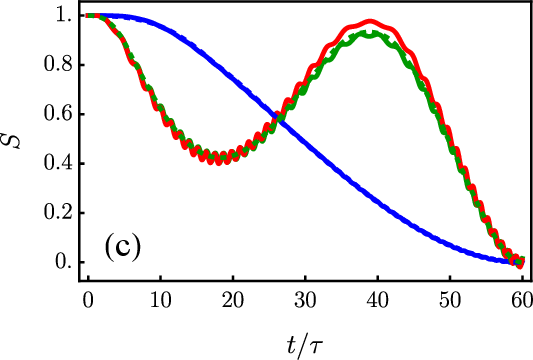}
\caption{Optimisation of the control: (a) Time-dependent angle $\phi(t)$, (b) trajectory of the piston $a(t)$ and velocity of the piston $v(t)$ (inset), (c) magnetization $S(t)$, everything with respect to time. In all figures: Reference control: 
evolution 
using stationary pressure $P_{\rm st}(a(t),\phi(t))$ (blue, dashed line), using exact pressure $P(t)$ (blue, solid line). Approximated optimal control with $c_{0,1}=0.276$ and $c_{0,2}=0.049$: 
evolution using approximated pressure $P_{\rm st}(a(t),\phi(t))$ (green, dashed line), using exact pressure $P(t)$ (green, solid line). Optimised control with $c_{1}=0.275$ and $c_{2}=0.025$: 
evolution using exact pressure $P(t)$ (red, solid line). $t_{f}/\tau=60$. The blue lines in (c) are on top of each other.
\label{fig5:examples}}
\end{figure}

\begin{figure}[t]
{\includegraphics[width=0.85\columnwidth]{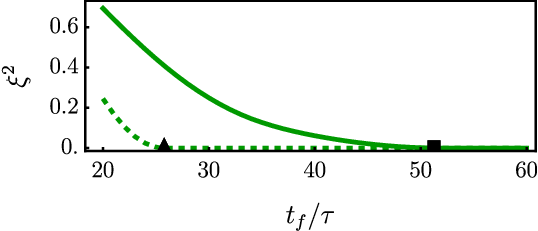}}
\caption{First-step optimisation: Error $\xi^{2}$ evaluated based on stationary pressure $P_{\rm st}$: 
piston mass $M/(Nm)=250$ (dashed lines), $M/(Nm)=1000$ (solid lines); the black triangle resp. box indicate the minimal required $t_{f,c}$ resulting in an error $\xi^{2} < 10^{-4}$ for the corresponding piston mass.
\label{fig6:e2}}
\end{figure}

To achieve this goal, we assume the time-dependent control function $\phi(t)$ given by
\begin{equation}
\phi(t)=\sum_{n=0}^{n_{\max}} l_{n} t^{n},
\end{equation}
where $l_{n}$ are the coefficients of polynomial function. 
The control function satisfies the following conditions augmenting Eq. \eqref{boundary1}:
\begin{eqnarray}
\label{eq:c1c2}
 \quad \phi(t_{f}/3)= \pi c_{1}, \quad \phi(2t_{f}/3)=\pi c_{2},
\end{eqnarray}
where $c_{1}$ and $c_{2}$ are control parameters. 
Thus, to satisfy the six conditions  in Eqs. \eqref{boundary1} and \eqref{eq:c1c2}, we use 5th-order polynomial function for interpolating.
The goal is to find parameters $c_{1}$ and $c_{2}$ which minimise this error function $\xi^{2}$
to achieve the goal of $\xi^{2} < 10^{-4}.$

Our optimisation strategy is motivated by the finding of the last subsection, namely that we achieve a good approximated piston trajectory by approximating the exact BEC pressure 
by the stationary pressure $P_{\rm st}(a(t),\phi(t))$. 
In detail, our optimisation strategy consists of two steps. In the first step, we will use this approximation, i.e. approximate the exact pressure by the stationary pressure $P_{\rm st}(a(t),\phi(t))$,
and we will then find values of control parameters $c_{1}$ and $c_{2}$ which minimised the error $\xi^{2}$.
It is important to underline that using the approximated $P_{\rm st}(a(t),\phi(t))$ makes it very easy to compute the corresponding error $\xi^{2}$. The reason is that $P_{\rm st}(a(t),\phi(t))$ has to be calculated numerically
only once for all final times and the whole optimisation. For a specific final time $t_{f}$, only a numerical solution of the time-dependent Newton equation of the piston, Eq. \eqref{approx} is required;
no solution of the GP equation is required for Eqs. (\ref{coupled_GPE}). Let us call the resulting control parameters of this first step $c_{0,1}$ and $c_{0,2}$.
We have calculated the optimal $c_{0,1}$ and $c_{0,2}$ for different final times and the achieved minimal error $\xi^{2}$ is plotted in Fig. \ref{fig6:e2}.
In this figure, we also consider two different piston masses, $M/(Nm)=250$ (dashed green line) and $M/(Nm)=1000$ (solid green line).  
It is interesting to note that we can see that there is a kind of "quantum speed limit" \cite{Deffner_2017} here, in the sense that a very small error $\xi^{2} < 10^{-4}$ can only be achieved for some minimal time $t_{f,c}$ onward. These minimal times for the two different masses are $t_{f,c}/\tau\approx 25.8$ and $t_{f,c}/\tau \approx 51.3$ for piston masses $M/(Nm)=250$ and $M/(Nm)=1000$, respectively. As one would expect the critical minimal final time scales approximately as $\sqrt{M},$ being of the order of piston oscillation period $\sim\sqrt{M/k}$ near the minimum of the potential $E_{\rm gs}(a,\Delta)+ka^{2}/2.$ We notice that in this hybrid system 
the speed limit is determined by dynamics of its classical component in the potential produced by the
quantum one.

In the next step, we will now consider the exact solutions without doing the approximation of a stationary pressure.
The key idea of the second step of our optimisation strategy is to use the parameters $c_{0,1}$ and $c_{0,2}$ resulting from the above described first step,as the starting point
for further optimisation, however now based on the exact pressure. In detail, the goal is to find optimal parameters $c_{1}, c_{2}$ which lead to an error $\xi^{2} < 10^{-4}$ in the exact quantum calculation using the exact ${\bm\Psi}-$determined pressure $P(t)$ instead of the 
approximated pressure $P_{\rm st}(a(t),\phi(t)).$

We will consider again the example shown in Fig. \ref{fig5:examples} with final time $t_{f}/\tau=60$ and a piston mass $M/(Nm)=1000$ for the second step of our optimisation.
Using the approximated $P_{\rm st}(a(t),\phi(t))$, we can find $c_{0,1}=0.276$ and $c_{0,2}=0.049$ from the first optimisation (based on the approximated $P_{\rm st}(a(t),\phi(t))$). 
The exact piston trajectory with these parameters is now shown in Fig. \ref{fig5:examples}(b) as the solid green line. Using the exact pressure, we achieve an error of $\xi^{2}=0.012$.
It turns out that the starting point is so close to the minimum using the exact $P(t)$, that the optimisation can be done just by considering all values in the close neighbourhood and looking for the minimum in a kind of "poor man approach". This results in values $c_{1}=0.275$ and $c_{2} = 0.025$ resulting in an error 
$\xi^{2}< 10^{-4}$ in the exact case. We present this optimal trajectory $\phi(t)/\pi$ as a red line in Fig. \ref{fig5:examples}(a) and the resulting piston trajectory as a red line in Fig. \ref{fig5:examples}(b).
It shows that the final length $a(t_{f})$ is close to the target equilibrium length $a_{T}$ in Fig. \ref{fig5:examples}(b) for the red line, even with using the exact pressure $P(t)$. 

The time-dependent magnetisation for the different cases is compared in Fig. \ref{fig5:examples}(c). We see that the average value of the exact magnetisation is very well approximated by the magnetisation for the simplified approximation corresponding to the stationary approximation. There are fluctuations in the exact magnetisation which cannot be seen in the approximation. It is also interesting to note that these fluctuations decrease with increasing final time $t_{f}$.

Let us come back to the mechanical work done by the piston and let us consider the example of the fully optimised control function for $t_f/\tau = 60$.
We are interested in the different components of this work, see Eq. \eqref{workcomponents}.
As the piston position is decreasing and the BEC is compressed the work must be negative. The total work done by piston is $W_{p}/\epsilon = -0.219$. 
The non-stationary contribution is $\delta W_{p}/\epsilon=0.001$, 
it is very small which is expected as we have seen above already that the exact pressure $P(t)$ can be well approximated the the stationary pressure. The change of the stationary energy of the BEC in this case is ${\mathcal E}_{\rm st} = E_{\rm st}(a(t_f),\pi/2) - E_{\rm st}(a(0),0)= -0.606 \,\epsilon$, 
i.e. the stationary system energy is decreasing by the process. The "stationary"
contribution to the work related to the control 
and change in $\phi$ is $W_{\phi,{\rm st}}/\epsilon=-0.826$ in this case.

\section{Conclusion}
\label{sec:4}
We studied the quantum control of a quantum-classical hybrid system made of a Rabi-coupled 
Bose-Einstein condensate with pseudospin $1/2$ interacting with a
classical heavy oscillator acting as a piston and producing a \textit{classical} 
mechanical work. The sufficiently large piston mass 
of the order of $10^{3}$ times the mass of the condensate consisting of $N\sim 10^{5}$ Rb atoms, 
can be achieved by using a low-density plastic particle of 
the radius of the order of 1 $\mu{\rm m}$, typical for manipulation with optical tweezers \cite{Ashkin1971,Ashkin1986}.
Notice that the mass of $\sim 10^{5}$ Rb atoms of mass $m\approx 1.5\times 10^{-22}$ g 
is of the order of $\sim 10^{-17}$ g and the piston mass 
corresponding to $M\sim 10^{3}mN$ is of the order of 
$\sim 10^{-14}$ g. For the BEC radius $\sim 0.5\mbox{ }\mu\mbox{m},$
this corresponds to the density of the order of $10^{-2}-10^{-1}$ 
g/cm$^{3}$  typical for a low-density plastic material. 
To provide an estimate of the classical work, we estimate the pressure as $N\pi^{2}\hbar^{2}/(ma^{3})$
and obtain for $N\sim 10^{5},$ $a\sim 10\mbox{ }\mu\mbox{m}$ its value 
$\sim 10^{-7}\mbox{ K}/\mu\mbox{m}$ with the corresponding classical work of the order 
of $10^{-7}\mbox{ K}$ per 1 $\mu\mbox{m}$ displacement. 

We demonstrated that by choosing the 
appropriate time dependence of the direction of the Rabi field in the condensate
not possessing the Manakov's rotational symmetry of the self-interaction, that is having 
different same- and opposite- spins interactions, one can control velocity and position of the 
piston and produced controllable mechanical work. This work is produced due 
to the fact that the Rabi field redistributes the self-interacting spin components 
of the condensate and modifies the pressure as a result. In numerical 
analysis of the optimal dynamical control we proposed and applied a two step process 
based on an approximation of the pressure function by its stationary value and subsequent 
corresponding correction based on the non-stationary behaviour.

In future works, it would be of interest to study, based on the model of this paper, 
full quantum heat engines with classical pistons and various 
quantum working bodies. Another interesting extension of this work 
is generalization of these approaches for a small mass quantum piston and 
studying the corresponding quantum effects.
\vspace{-0.25cm}

 \acknowledgements{
\vspace{-0.2cm}
 We are grateful to D. Rea for the fruitful discussion
and careful reading of the manuscript. 
 J.L. and A.R acknowledge that this publication has emanated from research supported in part by a research grant from Science Foundation Ireland (SFI) under Grant Number 19/FFP/6951.   
 The work of E.S. is financially supported through the Grant PGC2018-101355-B-I00 funded by MCIN/AEI/10.13039/501100011033 
 and by ERDF ``A way of making Europe'', and by the Basque Government through Grant No. IT986-16.}


\begin{thebibliography}{38}%
\makeatletter
\providecommand \@ifxundefined [1]{%
 \@ifx{#1\undefined}
}%
\providecommand \@ifnum [1]{%
 \ifnum #1\expandafter \@firstoftwo
 \else \expandafter \@secondoftwo
 \fi
}%
\providecommand \@ifx [1]{%
 \ifx #1\expandafter \@firstoftwo
 \else \expandafter \@secondoftwo
 \fi
}%
\providecommand \natexlab [1]{#1}%
\providecommand \enquote  [1]{``#1''}%
\providecommand \bibnamefont  [1]{#1}%
\providecommand \bibfnamefont [1]{#1}%
\providecommand \citenamefont [1]{#1}%
\providecommand \href@noop [0]{\@secondoftwo}%
\providecommand \href [0]{\begingroup \@sanitize@url \@href}%
\providecommand \@href[1]{\@@startlink{#1}\@@href}%
\providecommand \@@href[1]{\endgroup#1\@@endlink}%
\providecommand \@sanitize@url [0]{\catcode `\\12\catcode `\$12\catcode
  `\&12\catcode `\#12\catcode `\^12\catcode `\_12\catcode `\%12\relax}%
\providecommand \@@startlink[1]{}%
\providecommand \@@endlink[0]{}%
\providecommand \url  [0]{\begingroup\@sanitize@url \@url }%
\providecommand \@url [1]{\endgroup\@href {#1}{\urlprefix }}%
\providecommand \urlprefix  [0]{URL }%
\providecommand \Eprint [0]{\href }%
\providecommand \doibase [0]{https://doi.org/}%
\providecommand \selectlanguage [0]{\@gobble}%
\providecommand \bibinfo  [0]{\@secondoftwo}%
\providecommand \bibfield  [0]{\@secondoftwo}%
\providecommand \translation [1]{[#1]}%
\providecommand \BibitemOpen [0]{}%
\providecommand \bibitemStop [0]{}%
\providecommand \bibitemNoStop [0]{.\EOS\space}%
\providecommand \EOS [0]{\spacefactor3000\relax}%
\providecommand \BibitemShut  [1]{\csname bibitem#1\endcsname}%
\let\auto@bib@innerbib\@empty
\bibitem [{\citenamefont {Blickle}\ and\ \citenamefont
  {Bechinger}(2012)}]{Blickle2012}%
  \BibitemOpen
  \bibfield  {author} {\bibinfo {author} {\bibfnamefont {V.}~\bibnamefont
  {Blickle}}\ and\ \bibinfo {author} {\bibfnamefont {C.}~\bibnamefont
  {Bechinger}},\ }\href {https://doi.org/10.1038/nphys2163} {\bibfield
  {journal} {\bibinfo  {journal} {Nature Physics}\ }\textbf {\bibinfo {volume}
  {8}},\ \bibinfo {pages} {143} (\bibinfo {year} {2012})}\BibitemShut {NoStop}%
\bibitem [{\citenamefont {Goold}\ \emph {et~al.}(2016)\citenamefont {Goold},
  \citenamefont {Huber}, \citenamefont {Riera}, \citenamefont {del Rio},\ and\
  \citenamefont {Skrzypczyk}}]{Goold_2016}%
  \BibitemOpen
  \bibfield  {author} {\bibinfo {author} {\bibfnamefont {J.}~\bibnamefont
  {Goold}}, \bibinfo {author} {\bibfnamefont {M.}~\bibnamefont {Huber}},
  \bibinfo {author} {\bibfnamefont {A.}~\bibnamefont {Riera}}, \bibinfo
  {author} {\bibfnamefont {L.}~\bibnamefont {del Rio}},\ and\ \bibinfo {author}
  {\bibfnamefont {P.}~\bibnamefont {Skrzypczyk}},\ }\href
  {https://doi.org/10.1088/1751-8113/49/14/143001} {\bibfield  {journal}
  {\bibinfo  {journal} {Journal of Physics A: Mathematical and Theoretical}\
  }\textbf {\bibinfo {volume} {49}},\ \bibinfo {pages} {143001} (\bibinfo
  {year} {2016})}\BibitemShut {NoStop}%
\bibitem [{\citenamefont {Deffner}\ and\ \citenamefont
  {Campbell}(2019)}]{Stevebook}%
  \BibitemOpen
  \bibfield  {author} {\bibinfo {author} {\bibfnamefont {S.}~\bibnamefont
  {Deffner}}\ and\ \bibinfo {author} {\bibfnamefont {S.}~\bibnamefont
  {Campbell}},\ }\href {https://doi.org/10.1088/2053-2571/ab21c6} {\emph
  {\bibinfo {title} {Quantum Thermodynamics}}},\ 2053-2571\ (\bibinfo
  {publisher} {Morgan \& Claypool Publishers},\ \bibinfo {year}
  {2019})\BibitemShut {NoStop}%
\bibitem [{\citenamefont {Callen}(1991)}]{callen_textbook}%
  \BibitemOpen
  \bibfield  {author} {\bibinfo {author} {\bibfnamefont {H.}~\bibnamefont
  {Callen}},\ }\href {https://books.google.ie/books?id=m_39DwAAQBAJ} {\emph
  {\bibinfo {title} {Thermodynamics and an Introduction to Thermostatistics}}}\
  (\bibinfo  {publisher} {Wiley},\ \bibinfo {year} {1991})\BibitemShut
  {NoStop}%
\bibitem [{\citenamefont {Abah}\ \emph {et~al.}(2012)\citenamefont {Abah},
  \citenamefont {Ro\ss{}nagel}, \citenamefont {Jacob}, \citenamefont {Deffner},
  \citenamefont {Schmidt-Kaler}, \citenamefont {Singer},\ and\ \citenamefont
  {Lutz}}]{Lutz2012}%
  \BibitemOpen
  \bibfield  {author} {\bibinfo {author} {\bibfnamefont {O.}~\bibnamefont
  {Abah}}, \bibinfo {author} {\bibfnamefont {J.}~\bibnamefont {Ro\ss{}nagel}},
  \bibinfo {author} {\bibfnamefont {G.}~\bibnamefont {Jacob}}, \bibinfo
  {author} {\bibfnamefont {S.}~\bibnamefont {Deffner}}, \bibinfo {author}
  {\bibfnamefont {F.}~\bibnamefont {Schmidt-Kaler}}, \bibinfo {author}
  {\bibfnamefont {K.}~\bibnamefont {Singer}},\ and\ \bibinfo {author}
  {\bibfnamefont {E.}~\bibnamefont {Lutz}},\ }\href
  {https://doi.org/10.1103/PhysRevLett.109.203006} {\bibfield  {journal}
  {\bibinfo  {journal} {Phys. Rev. Lett.}\ }\textbf {\bibinfo {volume} {109}},\
  \bibinfo {pages} {203006} (\bibinfo {year} {2012})}\BibitemShut {NoStop}%
\bibitem [{\citenamefont {Roßnagel}\ \emph {et~al.}(2016)\citenamefont
  {Roßnagel}, \citenamefont {Dawkins}, \citenamefont {Tolazzi}, \citenamefont
  {Abah}, \citenamefont {Lutz}, \citenamefont {Schmidt-Kaler},\ and\
  \citenamefont {Singer}}]{Kilian2016}%
  \BibitemOpen
  \bibfield  {author} {\bibinfo {author} {\bibfnamefont {J.}~\bibnamefont
  {Roßnagel}}, \bibinfo {author} {\bibfnamefont {S.~T.}\ \bibnamefont
  {Dawkins}}, \bibinfo {author} {\bibfnamefont {K.~N.}\ \bibnamefont
  {Tolazzi}}, \bibinfo {author} {\bibfnamefont {O.}~\bibnamefont {Abah}},
  \bibinfo {author} {\bibfnamefont {E.}~\bibnamefont {Lutz}}, \bibinfo {author}
  {\bibfnamefont {F.}~\bibnamefont {Schmidt-Kaler}},\ and\ \bibinfo {author}
  {\bibfnamefont {K.}~\bibnamefont {Singer}},\ }\href
  {https://doi.org/10.1126/science.aad6320} {\bibfield  {journal} {\bibinfo
  {journal} {Science}\ }\textbf {\bibinfo {volume} {352}},\ \bibinfo {pages}
  {325} (\bibinfo {year} {2016})}\BibitemShut {NoStop}%
\bibitem [{\citenamefont {Bouton}\ \emph {et~al.}(2021)\citenamefont {Bouton},
  \citenamefont {Nettersheim}, \citenamefont {Adam}, \citenamefont {Lutz},\
  and\ \citenamefont {Widera}}]{Widera21}%
  \BibitemOpen
  \bibfield  {author} {\bibinfo {author} {\bibfnamefont {Q.}~\bibnamefont
  {Bouton}}, \bibinfo {author} {\bibfnamefont {S.}~\bibnamefont {Nettersheim},
  \bibfnamefont {Jensand~Burgardt}}, \bibinfo {author} {\bibfnamefont
  {D.}~\bibnamefont {Adam}}, \bibinfo {author} {\bibfnamefont {E.}~\bibnamefont
  {Lutz}},\ and\ \bibinfo {author} {\bibfnamefont {A.}~\bibnamefont {Widera}},\
  }\href {https://doi.org/10.1038/s41467-021-22222-z} {\bibfield  {journal}
  {\bibinfo  {journal} {Nature Communications}\ }\textbf {\bibinfo {volume}
  {12}},\ \bibinfo {pages} {2063} (\bibinfo {year} {2021})}\BibitemShut
  {NoStop}%
\bibitem [{\citenamefont {Sothmann}\ \emph {et~al.}(2014)\citenamefont
  {Sothmann}, \citenamefont {S{\'{a}}nchez},\ and\ \citenamefont
  {Jordan}}]{Sothmann_2014}%
  \BibitemOpen
  \bibfield  {author} {\bibinfo {author} {\bibfnamefont {B.}~\bibnamefont
  {Sothmann}}, \bibinfo {author} {\bibfnamefont {R.}~\bibnamefont
  {S{\'{a}}nchez}},\ and\ \bibinfo {author} {\bibfnamefont {A.~N.}\
  \bibnamefont {Jordan}},\ }\href
  {https://doi.org/10.1088/0957-4484/26/3/032001} {\bibfield  {journal}
  {\bibinfo  {journal} {Nanotechnology}\ }\textbf {\bibinfo {volume} {26}},\
  \bibinfo {pages} {032001} (\bibinfo {year} {2014})}\BibitemShut {NoStop}%
\bibitem [{\citenamefont {Ono}\ \emph {et~al.}(2020)\citenamefont {Ono},
  \citenamefont {Shevchenko}, \citenamefont {Mori}, \citenamefont {Moriyama},\
  and\ \citenamefont {Nori}}]{Franco2021}%
  \BibitemOpen
  \bibfield  {author} {\bibinfo {author} {\bibfnamefont {K.}~\bibnamefont
  {Ono}}, \bibinfo {author} {\bibfnamefont {S.~N.}\ \bibnamefont {Shevchenko}},
  \bibinfo {author} {\bibfnamefont {T.}~\bibnamefont {Mori}}, \bibinfo {author}
  {\bibfnamefont {S.}~\bibnamefont {Moriyama}},\ and\ \bibinfo {author}
  {\bibfnamefont {F.}~\bibnamefont {Nori}},\ }\href
  {https://doi.org/10.1103/PhysRevLett.125.166802} {\bibfield  {journal}
  {\bibinfo  {journal} {Phys. Rev. Lett.}\ }\textbf {\bibinfo {volume} {125}},\
  \bibinfo {pages} {166802} (\bibinfo {year} {2020})}\BibitemShut {NoStop}%
\bibitem [{\citenamefont {Zhang}\ \emph {et~al.}(2014)\citenamefont {Zhang},
  \citenamefont {Bariani},\ and\ \citenamefont {Meystre}}]{Pierre2014}%
  \BibitemOpen
  \bibfield  {author} {\bibinfo {author} {\bibfnamefont {K.}~\bibnamefont
  {Zhang}}, \bibinfo {author} {\bibfnamefont {F.}~\bibnamefont {Bariani}},\
  and\ \bibinfo {author} {\bibfnamefont {P.}~\bibnamefont {Meystre}},\ }\href
  {https://doi.org/10.1103/PhysRevLett.112.150602} {\bibfield  {journal}
  {\bibinfo  {journal} {Phys. Rev. Lett.}\ }\textbf {\bibinfo {volume} {112}},\
  \bibinfo {pages} {150602} (\bibinfo {year} {2014})}\BibitemShut {NoStop}%
\bibitem [{\citenamefont {Bergenfeldt}\ \emph {et~al.}(2014)\citenamefont
  {Bergenfeldt}, \citenamefont {Samuelsson}, \citenamefont {Sothmann},
  \citenamefont {Flindt},\ and\ \citenamefont {B\"uttiker}}]{Markus2014}%
  \BibitemOpen
  \bibfield  {author} {\bibinfo {author} {\bibfnamefont {C.}~\bibnamefont
  {Bergenfeldt}}, \bibinfo {author} {\bibfnamefont {P.}~\bibnamefont
  {Samuelsson}}, \bibinfo {author} {\bibfnamefont {B.}~\bibnamefont
  {Sothmann}}, \bibinfo {author} {\bibfnamefont {C.}~\bibnamefont {Flindt}},\
  and\ \bibinfo {author} {\bibfnamefont {M.}~\bibnamefont {B\"uttiker}},\
  }\href {https://doi.org/10.1103/PhysRevLett.112.076803} {\bibfield  {journal}
  {\bibinfo  {journal} {Phys. Rev. Lett.}\ }\textbf {\bibinfo {volume} {112}},\
  \bibinfo {pages} {076803} (\bibinfo {year} {2014})}\BibitemShut {NoStop}%
\bibitem [{\citenamefont {Elouard}\ \emph {et~al.}(2015)\citenamefont
  {Elouard}, \citenamefont {Richard},\ and\ \citenamefont
  {Auff{\`{e}}ves}}]{Elouard_2015}%
  \BibitemOpen
  \bibfield  {author} {\bibinfo {author} {\bibfnamefont {C.}~\bibnamefont
  {Elouard}}, \bibinfo {author} {\bibfnamefont {M.}~\bibnamefont {Richard}},\
  and\ \bibinfo {author} {\bibfnamefont {A.}~\bibnamefont {Auff{\`{e}}ves}},\
  }\href {https://doi.org/10.1088/1367-2630/17/5/055018} {\bibfield  {journal}
  {\bibinfo  {journal} {New Journal of Physics}\ }\textbf {\bibinfo {volume}
  {17}},\ \bibinfo {pages} {055018} (\bibinfo {year} {2015})}\BibitemShut
  {NoStop}%
\bibitem [{\citenamefont {Hugel}\ \emph {et~al.}(2002)\citenamefont {Hugel},
  \citenamefont {Holland}, \citenamefont {Cattani}, \citenamefont {Moroder},
  \citenamefont {Seitz},\ and\ \citenamefont {Gaub}}]{Gaub2002}%
  \BibitemOpen
  \bibfield  {author} {\bibinfo {author} {\bibfnamefont {T.}~\bibnamefont
  {Hugel}}, \bibinfo {author} {\bibfnamefont {N.~B.}\ \bibnamefont {Holland}},
  \bibinfo {author} {\bibfnamefont {A.}~\bibnamefont {Cattani}}, \bibinfo
  {author} {\bibfnamefont {L.}~\bibnamefont {Moroder}}, \bibinfo {author}
  {\bibfnamefont {M.}~\bibnamefont {Seitz}},\ and\ \bibinfo {author}
  {\bibfnamefont {H.~E.}\ \bibnamefont {Gaub}},\ }\href
  {https://doi.org/10.1126/science.1069856} {\bibfield  {journal} {\bibinfo
  {journal} {Science}\ }\textbf {\bibinfo {volume} {296}},\ \bibinfo {pages}
  {1103} (\bibinfo {year} {2002})}\BibitemShut {NoStop}%
\bibitem [{\citenamefont {Charalambous}\ \emph {et~al.}(2019)\citenamefont
  {Charalambous}, \citenamefont {Garcia-March}, \citenamefont {Mehboudi},\ and\
  \citenamefont {Lewenstein}}]{Charalambous_2019}%
  \BibitemOpen
  \bibfield  {author} {\bibinfo {author} {\bibfnamefont {C.}~\bibnamefont
  {Charalambous}}, \bibinfo {author} {\bibfnamefont {M.~A.}\ \bibnamefont
  {Garcia-March}}, \bibinfo {author} {\bibfnamefont {M.}~\bibnamefont
  {Mehboudi}},\ and\ \bibinfo {author} {\bibfnamefont {M.}~\bibnamefont
  {Lewenstein}},\ }\href {https://doi.org/10.1088/1367-2630/ab3832} {\bibfield
  {journal} {\bibinfo  {journal} {New Journal of Physics}\ }\textbf {\bibinfo
  {volume} {21}},\ \bibinfo {pages} {083037} (\bibinfo {year}
  {2019})}\BibitemShut {NoStop}%
\bibitem [{\citenamefont {Myers}\ \emph {et~al.}(2022)\citenamefont {Myers},
  \citenamefont {Pe{\~{n}}a}, \citenamefont {Negrete}, \citenamefont {Vargas},
  \citenamefont {Chiara},\ and\ \citenamefont {Deffner}}]{Myers_2022}%
  \BibitemOpen
  \bibfield  {author} {\bibinfo {author} {\bibfnamefont {N.~M.}\ \bibnamefont
  {Myers}}, \bibinfo {author} {\bibfnamefont {F.~J.}\ \bibnamefont
  {Pe{\~{n}}a}}, \bibinfo {author} {\bibfnamefont {O.}~\bibnamefont {Negrete}},
  \bibinfo {author} {\bibfnamefont {P.}~\bibnamefont {Vargas}}, \bibinfo
  {author} {\bibfnamefont {G.~D.}\ \bibnamefont {Chiara}},\ and\ \bibinfo
  {author} {\bibfnamefont {S.}~\bibnamefont {Deffner}},\ }\href
  {https://doi.org/10.1088/1367-2630/ac47cc} {\bibfield  {journal} {\bibinfo
  {journal} {New Journal of Physics}\ }\textbf {\bibinfo {volume} {24}},\
  \bibinfo {pages} {025001} (\bibinfo {year} {2022})}\BibitemShut {NoStop}%
\bibitem [{\citenamefont {Li}\ \emph {et~al.}(2018)\citenamefont {Li},
  \citenamefont {Fogarty}, \citenamefont {Campbell}, \citenamefont {Chen},\
  and\ \citenamefont {Busch}}]{Li2018}%
  \BibitemOpen
  \bibfield  {author} {\bibinfo {author} {\bibfnamefont {J.}~\bibnamefont
  {Li}}, \bibinfo {author} {\bibfnamefont {T.}~\bibnamefont {Fogarty}},
  \bibinfo {author} {\bibfnamefont {S.}~\bibnamefont {Campbell}}, \bibinfo
  {author} {\bibfnamefont {X.}~\bibnamefont {Chen}},\ and\ \bibinfo {author}
  {\bibfnamefont {T.}~\bibnamefont {Busch}},\ }\href
  {https://doi.org/10.1088/1367-2630/aa9cd8} {\bibfield  {journal} {\bibinfo
  {journal} {New Journal of Physics}\ }\textbf {\bibinfo {volume} {20}},\
  \bibinfo {pages} {015005} (\bibinfo {year} {2018})}\BibitemShut {NoStop}%
\bibitem [{\citenamefont {Niedenzu}\ \emph {et~al.}(2019)\citenamefont
  {Niedenzu}, \citenamefont {Mazets}, \citenamefont {Kurizki},\ and\
  \citenamefont {Jendrzejewski}}]{Niedenzu2019}%
  \BibitemOpen
  \bibfield  {author} {\bibinfo {author} {\bibfnamefont {W.}~\bibnamefont
  {Niedenzu}}, \bibinfo {author} {\bibfnamefont {I.}~\bibnamefont {Mazets}},
  \bibinfo {author} {\bibfnamefont {G.}~\bibnamefont {Kurizki}},\ and\ \bibinfo
  {author} {\bibfnamefont {F.}~\bibnamefont {Jendrzejewski}},\ }\href
  {https://doi.org/10.22331/q-2019-06-28-155} {\bibfield  {journal} {\bibinfo
  {journal} {{Quantum}}\ }\textbf {\bibinfo {volume} {3}},\ \bibinfo {pages}
  {155} (\bibinfo {year} {2019})}\BibitemShut {NoStop}%
\bibitem [{\citenamefont {Li}\ \emph {et~al.}(2022)\citenamefont {Li},
  \citenamefont {Sherman},\ and\ \citenamefont {Ruschhaupt}}]{Jing22}%
  \BibitemOpen
  \bibfield  {author} {\bibinfo {author} {\bibfnamefont {J.}~\bibnamefont
  {Li}}, \bibinfo {author} {\bibfnamefont {E.~Y.}\ \bibnamefont {Sherman}},\
  and\ \bibinfo {author} {\bibfnamefont {A.}~\bibnamefont {Ruschhaupt}},\
  }\href {https://doi.org/10.1103/PhysRevA.106.L030201} {\bibfield  {journal}
  {\bibinfo  {journal} {Phys. Rev. A}\ }\textbf {\bibinfo {volume} {106}},\
  \bibinfo {pages} {L030201} (\bibinfo {year} {2022})}\BibitemShut {NoStop}%
\bibitem [{\citenamefont {Koch}\ \emph {et~al.}(2023)\citenamefont {Koch},
  \citenamefont {Menon}, \citenamefont {Cuestas}, \citenamefont {Barbosa},
  \citenamefont {Lutz}, \citenamefont {Fogarty}, \citenamefont {Busch},\ and\
  \citenamefont {Widera}}]{Koch2022}%
  \BibitemOpen
  \bibfield  {author} {\bibinfo {author} {\bibfnamefont {J.}~\bibnamefont
  {Koch}}, \bibinfo {author} {\bibfnamefont {K.}~\bibnamefont {Menon}},
  \bibinfo {author} {\bibfnamefont {E.}~\bibnamefont {Cuestas}}, \bibinfo
  {author} {\bibfnamefont {S.}~\bibnamefont {Barbosa}}, \bibinfo {author}
  {\bibfnamefont {E.}~\bibnamefont {Lutz}}, \bibinfo {author} {\bibfnamefont
  {T.}~\bibnamefont {Fogarty}}, \bibinfo {author} {\bibfnamefont
  {T.}~\bibnamefont {Busch}},\ and\ \bibinfo {author} {\bibfnamefont
  {A.}~\bibnamefont {Widera}},\ }\href
  {https://doi.org/10.1038/s41586-023-06469-8} {\bibfield  {journal} {\bibinfo
  {journal} {Nature}\ }\textbf {\bibinfo {volume} {621}},\ \bibinfo {pages}
  {723} (\bibinfo {year} {2023})}\BibitemShut {NoStop}%
\bibitem [{\citenamefont {Torrontegui}\ \emph {et~al.}(2013)\citenamefont
  {Torrontegui}, \citenamefont {Ibáñez}, \citenamefont {Martínez-Garaot},
  \citenamefont {Modugno}, \citenamefont {{del Campo}}, \citenamefont
  {Guéry-Odelin}, \citenamefont {Ruschhaupt}, \citenamefont {Chen},\ and\
  \citenamefont {Muga}}]{TORRONTEGUI2013117}%
  \BibitemOpen
  \bibfield  {author} {\bibinfo {author} {\bibfnamefont {E.}~\bibnamefont
  {Torrontegui}}, \bibinfo {author} {\bibfnamefont {S.}~\bibnamefont
  {Ibáñez}}, \bibinfo {author} {\bibfnamefont {S.}~\bibnamefont
  {Martínez-Garaot}}, \bibinfo {author} {\bibfnamefont {M.}~\bibnamefont
  {Modugno}}, \bibinfo {author} {\bibfnamefont {A.}~\bibnamefont {{del
  Campo}}}, \bibinfo {author} {\bibfnamefont {D.}~\bibnamefont
  {Guéry-Odelin}}, \bibinfo {author} {\bibfnamefont {A.}~\bibnamefont
  {Ruschhaupt}}, \bibinfo {author} {\bibfnamefont {X.}~\bibnamefont {Chen}},\
  and\ \bibinfo {author} {\bibfnamefont {J.~G.}\ \bibnamefont {Muga}},\ }in\
  \href {https://doi.org/https://doi.org/10.1016/B978-0-12-408090-4.00002-5}
  {\emph {\bibinfo {booktitle} {Advances in Atomic, Molecular, and Optical
  Physics}}},\  Vol.~\bibinfo {volume} {62}, pp.\ \bibinfo {pages}
  {117--169}  (\bibinfo  {publisher}
  {Academic Press},\ \bibinfo {year} {2013}) \BibitemShut {NoStop}%
\bibitem [{\citenamefont {Gu\'ery-Odelin}\ \emph {et~al.}(2019)\citenamefont
  {Gu\'ery-Odelin}, \citenamefont {Ruschhaupt}, \citenamefont {Kiely},
  \citenamefont {Torrontegui}, \citenamefont {Mart\'{\i}nez-Garaot},\ and\
  \citenamefont {Muga}}]{STAreview}%
  \BibitemOpen
  \bibfield  {author} {\bibinfo {author} {\bibfnamefont {D.}~\bibnamefont
  {Gu\'ery-Odelin}}, \bibinfo {author} {\bibfnamefont {A.}~\bibnamefont
  {Ruschhaupt}}, \bibinfo {author} {\bibfnamefont {A.}~\bibnamefont {Kiely}},
  \bibinfo {author} {\bibfnamefont {E.}~\bibnamefont {Torrontegui}}, \bibinfo
  {author} {\bibfnamefont {S.}~\bibnamefont {Mart\'{\i}nez-Garaot}},\ and\
  \bibinfo {author} {\bibfnamefont {J.~G.}\ \bibnamefont {Muga}},\ }\href
  {https://doi.org/10.1103/RevModPhys.91.045001} {\bibfield  {journal}
  {\bibinfo  {journal} {Rev. Mod. Phys.}\ }\textbf {\bibinfo {volume} {91}},\
  \bibinfo {pages} {045001} (\bibinfo {year} {2019})}\BibitemShut {NoStop}%
\bibitem [{\citenamefont {Khaneja}\ \emph {et~al.}(2005)\citenamefont
  {Khaneja}, \citenamefont {Reiss}, \citenamefont {Kehlet}, \citenamefont
  {Schulte-Herbrüggen},\ and\ \citenamefont {Glaser}}]{Glaser2005}%
  \BibitemOpen
  \bibfield  {author} {\bibinfo {author} {\bibfnamefont {N.}~\bibnamefont
  {Khaneja}}, \bibinfo {author} {\bibfnamefont {T.}~\bibnamefont {Reiss}},
  \bibinfo {author} {\bibfnamefont {C.}~\bibnamefont {Kehlet}}, \bibinfo
  {author} {\bibfnamefont {T.}~\bibnamefont {Schulte-Herbrüggen}},\ and\
  \bibinfo {author} {\bibfnamefont {S.~J.}\ \bibnamefont {Glaser}},\ }\href
  {https://doi.org/https://doi.org/10.1016/j.jmr.2004.11.004} {\bibfield
  {journal} {\bibinfo  {journal} {Journal of Magnetic Resonance}\ }\textbf
  {\bibinfo {volume} {172}},\ \bibinfo {pages} {296} (\bibinfo {year}
  {2005})}\BibitemShut {NoStop}%
\bibitem [{\citenamefont {Doria}\ \emph {et~al.}(2011)\citenamefont {Doria},
  \citenamefont {Calarco},\ and\ \citenamefont {Montangero}}]{Montangero2011}%
  \BibitemOpen
  \bibfield  {author} {\bibinfo {author} {\bibfnamefont {P.}~\bibnamefont
  {Doria}}, \bibinfo {author} {\bibfnamefont {T.}~\bibnamefont {Calarco}},\
  and\ \bibinfo {author} {\bibfnamefont {S.}~\bibnamefont {Montangero}},\
  }\href {https://doi.org/10.1103/PhysRevLett.106.190501} {\bibfield  {journal}
  {\bibinfo  {journal} {Phys. Rev. Lett.}\ }\textbf {\bibinfo {volume} {106}},\
  \bibinfo {pages} {190501} (\bibinfo {year} {2011})}\BibitemShut {NoStop}%
\bibitem [{\citenamefont {Glaser}\ \emph {et~al.}(2015)\citenamefont {Glaser},
  \citenamefont {Boscain}, \citenamefont {Calarco}, \citenamefont {Koch},
  \citenamefont {K{\"o}ckenberger}, \citenamefont {Kosloff}, \citenamefont
  {Kuprov}, \citenamefont {Luy}, \citenamefont {Schirmer}, \citenamefont
  {Schulte-Herbr{\"u}ggen}, \citenamefont {Sugny},\ and\ \citenamefont
  {Wilhelm}}]{Glaser2015}%
  \BibitemOpen
  \bibfield  {author} {\bibinfo {author} {\bibfnamefont {S.~J.}\ \bibnamefont
  {Glaser}}, \bibinfo {author} {\bibfnamefont {U.}~\bibnamefont {Boscain}},
  \bibinfo {author} {\bibfnamefont {T.}~\bibnamefont {Calarco}}, \bibinfo
  {author} {\bibfnamefont {C.~P.}\ \bibnamefont {Koch}}, \bibinfo {author}
  {\bibfnamefont {W.}~\bibnamefont {K{\"o}ckenberger}}, \bibinfo {author}
  {\bibfnamefont {R.}~\bibnamefont {Kosloff}}, \bibinfo {author} {\bibfnamefont
  {I.}~\bibnamefont {Kuprov}}, \bibinfo {author} {\bibfnamefont
  {B.}~\bibnamefont {Luy}}, \bibinfo {author} {\bibfnamefont {S.}~\bibnamefont
  {Schirmer}}, \bibinfo {author} {\bibfnamefont {T.}~\bibnamefont
  {Schulte-Herbr{\"u}ggen}}, \bibinfo {author} {\bibfnamefont {D.}~\bibnamefont
  {Sugny}},\ and\ \bibinfo {author} {\bibfnamefont {F.~K.}\ \bibnamefont
  {Wilhelm}},\ }\href {https://doi.org/10.1140/epjd/e2015-60464-1} {\bibfield
  {journal} {\bibinfo  {journal} {The European Physical Journal D}\ }\textbf
  {\bibinfo {volume} {69}},\ \bibinfo {pages} {279} (\bibinfo {year}
  {2015})}\BibitemShut {NoStop}%
\bibitem [{\citenamefont {Poggi}\ \emph {et~al.}(2023)\citenamefont {Poggi},
  \citenamefont {Chiara}, \citenamefont {Campbell},\ and\ \citenamefont
  {Kiely}}]{Kiely2023}%
  \BibitemOpen
  \bibfield  {author} {\bibinfo {author} {\bibfnamefont {P.~M.}\ \bibnamefont
  {Poggi}}, \bibinfo {author} {\bibfnamefont {G.~D.}\ \bibnamefont {Chiara}},
  \bibinfo {author} {\bibfnamefont {S.}~\bibnamefont {Campbell}},\ and\
  \bibinfo {author} {\bibfnamefont {A.}~\bibnamefont {Kiely}},\ }\href@noop {}
  {\bibinfo {title} {Universally robust quantum control}} (\bibinfo {year}
  {2023}),\ \Eprint {https://arxiv.org/abs/2309.14437} {arXiv:2309.14437
  [quant-ph]} \BibitemShut {NoStop}%
\bibitem [{\citenamefont {Kieu}(2004)}]{KieuPRL04}%
  \BibitemOpen
  \bibfield  {author} {\bibinfo {author} {\bibfnamefont {T.~D.}\ \bibnamefont
  {Kieu}},\ }\href {https://doi.org/10.1103/PhysRevLett.93.140403} {\bibfield
  {journal} {\bibinfo  {journal} {Phys. Rev. Lett.}\ }\textbf {\bibinfo
  {volume} {93}},\ \bibinfo {pages} {140403} (\bibinfo {year}
  {2004})}\BibitemShut {NoStop}%
\bibitem [{\citenamefont {Talkner}\ \emph {et~al.}(2007)\citenamefont
  {Talkner}, \citenamefont {Lutz},\ and\ \citenamefont {H\"anggi}}]{Hanggi07}%
  \BibitemOpen
  \bibfield  {author} {\bibinfo {author} {\bibfnamefont {P.}~\bibnamefont
  {Talkner}}, \bibinfo {author} {\bibfnamefont {E.}~\bibnamefont {Lutz}},\ and\
  \bibinfo {author} {\bibfnamefont {P.}~\bibnamefont {H\"anggi}},\ }\href
  {https://doi.org/10.1103/PhysRevE.75.050102} {\bibfield  {journal} {\bibinfo
  {journal} {Phys. Rev. E}\ }\textbf {\bibinfo {volume} {75}},\ \bibinfo
  {pages} {050102} (\bibinfo {year} {2007})}\BibitemShut {NoStop}%
\bibitem [{\citenamefont {Campisi}\ \emph {et~al.}(2011)\citenamefont
  {Campisi}, \citenamefont {H\"anggi},\ and\ \citenamefont
  {Talkner}}]{TalknerRMP2011}%
  \BibitemOpen
  \bibfield  {author} {\bibinfo {author} {\bibfnamefont {M.}~\bibnamefont
  {Campisi}}, \bibinfo {author} {\bibfnamefont {P.}~\bibnamefont {H\"anggi}},\
  and\ \bibinfo {author} {\bibfnamefont {P.}~\bibnamefont {Talkner}},\ }\href
  {https://doi.org/10.1103/RevModPhys.83.771} {\bibfield  {journal} {\bibinfo
  {journal} {Rev. Mod. Phys.}\ }\textbf {\bibinfo {volume} {83}},\ \bibinfo
  {pages} {771} (\bibinfo {year} {2011})}\BibitemShut {NoStop}%
\bibitem [{\citenamefont {Verteletsky}\ and\ \citenamefont
  {M\o{}lmer}(2020)}]{Molmer2020}%
  \BibitemOpen
  \bibfield  {author} {\bibinfo {author} {\bibfnamefont {K.}~\bibnamefont
  {Verteletsky}}\ and\ \bibinfo {author} {\bibfnamefont {K.}~\bibnamefont
  {M\o{}lmer}},\ }\href {https://doi.org/10.1103/PhysRevA.101.010101}
  {\bibfield  {journal} {\bibinfo  {journal} {Phys. Rev. A}\ }\textbf {\bibinfo
  {volume} {101}},\ \bibinfo {pages} {010101} (\bibinfo {year}
  {2020})}\BibitemShut {NoStop}%
\bibitem [{\citenamefont {Gelbwaser-Klimovsky}\ and\ \citenamefont
  {Kurizki}(2014)}]{Kurizki14}%
  \BibitemOpen
  \bibfield  {author} {\bibinfo {author} {\bibfnamefont {D.}~\bibnamefont
  {Gelbwaser-Klimovsky}}\ and\ \bibinfo {author} {\bibfnamefont
  {G.}~\bibnamefont {Kurizki}},\ }\href
  {https://doi.org/10.1103/PhysRevE.90.022102} {\bibfield  {journal} {\bibinfo
  {journal} {Phys. Rev. E}\ }\textbf {\bibinfo {volume} {90}},\ \bibinfo
  {pages} {022102} (\bibinfo {year} {2014})}\BibitemShut {NoStop}%
\bibitem [{\citenamefont {Ashkin}\ and\ \citenamefont
  {Dziedzic}(1971)}]{Ashkin1971}%
  \BibitemOpen
  \bibfield  {author} {\bibinfo {author} {\bibfnamefont {A.}~\bibnamefont
  {Ashkin}}\ and\ \bibinfo {author} {\bibfnamefont {J.}~\bibnamefont
  {Dziedzic}},\ }\href@noop {} {\bibfield  {journal} {\bibinfo  {journal}
  {Applied Physics Letters}\ }\textbf {\bibinfo {volume} {19}},\ \bibinfo
  {pages} {283} (\bibinfo {year} {1971})}\BibitemShut {NoStop}%
\bibitem [{\citenamefont {Ashkin}\ \emph {et~al.}(1986)\citenamefont {Ashkin},
  \citenamefont {Dziedzic}, \citenamefont {Bjorkholm},\ and\ \citenamefont
  {Chu}}]{Ashkin1986}%
  \BibitemOpen
  \bibfield  {author} {\bibinfo {author} {\bibfnamefont {A.}~\bibnamefont
  {Ashkin}}, \bibinfo {author} {\bibfnamefont {J.}~\bibnamefont {Dziedzic}},
  \bibinfo {author} {\bibfnamefont {J.~E.}\ \bibnamefont {Bjorkholm}},\ and\
  \bibinfo {author} {\bibfnamefont {S.}~\bibnamefont {Chu}},\ }\href
  {https://doi.org/10.1364/OL.11.000288} {\bibfield  {journal} {\bibinfo
  {journal} {Optics Letters}\ }\textbf {\bibinfo {volume} {11}},\ \bibinfo
  {pages} {288} (\bibinfo {year} {1986})}\BibitemShut {NoStop}%
\bibitem [{\citenamefont {Courteille}\ \emph {et~al.}(1998)\citenamefont
  {Courteille}, \citenamefont {Freeland}, \citenamefont {Heinzen},
  \citenamefont {van Abeelen},\ and\ \citenamefont {Verhaar}}]{Feshbach98}%
  \BibitemOpen
  \bibfield  {author} {\bibinfo {author} {\bibfnamefont {P.}~\bibnamefont
  {Courteille}}, \bibinfo {author} {\bibfnamefont {R.~S.}\ \bibnamefont
  {Freeland}}, \bibinfo {author} {\bibfnamefont {D.~J.}\ \bibnamefont
  {Heinzen}}, \bibinfo {author} {\bibfnamefont {F.~A.}\ \bibnamefont {van
  Abeelen}},\ and\ \bibinfo {author} {\bibfnamefont {B.~J.}\ \bibnamefont
  {Verhaar}},\ }\href {https://doi.org/10.1103/PhysRevLett.81.69} {\bibfield
  {journal} {\bibinfo  {journal} {Phys. Rev. Lett.}\ }\textbf {\bibinfo
  {volume} {81}},\ \bibinfo {pages} {69} (\bibinfo {year} {1998})}\BibitemShut
  {NoStop}%
\bibitem [{\citenamefont {Manakov}(1974)}]{Manakov1974}%
  \BibitemOpen
  \bibfield  {author} {\bibinfo {author} {\bibfnamefont {S.~V.}\ \bibnamefont
  {Manakov}},\ }\href@noop {} {\bibfield  {journal} {\bibinfo  {journal} {Sov.
  Phys. JETP}\ }\textbf {\bibinfo {volume} {65}},\ \bibinfo {pages} {248}
  (\bibinfo {year} {1974})}\BibitemShut {NoStop}%
\bibitem [{\citenamefont {Mardonov}\ \emph {et~al.}(2019)\citenamefont
  {Mardonov}, \citenamefont {Modugno}, \citenamefont {Sherman},\ and\
  \citenamefont {Malomed}}]{Mardonov2019}%
  \BibitemOpen
  \bibfield  {author} {\bibinfo {author} {\bibfnamefont {S.}~\bibnamefont
  {Mardonov}}, \bibinfo {author} {\bibfnamefont {M.}~\bibnamefont {Modugno}},
  \bibinfo {author} {\bibfnamefont {E.~Y.}\ \bibnamefont {Sherman}},\ and\
  \bibinfo {author} {\bibfnamefont {B.~A.}\ \bibnamefont {Malomed}},\ }\href
  {https://doi.org/10.1364/OL.11.000288} {\bibfield  {journal} {\bibinfo
  {journal} {Phys. Rev. A}\ }\textbf {\bibinfo {volume} {99}},\ \bibinfo
  {pages} {013611} (\bibinfo {year} {2019})}\BibitemShut {NoStop}%
\bibitem [{\citenamefont {Eichmann}\ and\ \citenamefont
  {Anglin}(2021)}]{Eichmann2021}%
  \BibitemOpen
  \bibfield  {author} {\bibinfo {author} {\bibfnamefont {T.}~\bibnamefont
  {Eichmann}}\ and\ \bibinfo {author} {\bibfnamefont {J.~R.}\ \bibnamefont
  {Anglin}},\ }\href {https://doi.org/10.1103/PhysRevA.104.043317} {\bibfield
  {journal} {\bibinfo  {journal} {Phys. Rev. A}\ }\textbf {\bibinfo {volume}
  {104}},\ \bibinfo {pages} {043317} (\bibinfo {year} {2021})}\BibitemShut
  {NoStop}%
\bibitem [{\citenamefont {Sarkar}\ \emph {et~al.}(2023)\citenamefont {Sarkar},
  \citenamefont {Mishra}, \citenamefont {Muruganandam},\ and\ \citenamefont
  {Mishra}}]{Sarkar2023}%
  \BibitemOpen
  \bibfield  {author} {\bibinfo {author} {\bibfnamefont {S.~K.}\ \bibnamefont
  {Sarkar}}, \bibinfo {author} {\bibfnamefont {T.}~\bibnamefont {Mishra}},
  \bibinfo {author} {\bibfnamefont {P.}~\bibnamefont {Muruganandam}},\ and\
  \bibinfo {author} {\bibfnamefont {P.~K.}\ \bibnamefont {Mishra}},\ }\href
  {https://doi.org/10.1103/PhysRevA.107.053320} {\bibfield  {journal} {\bibinfo
   {journal} {Phys. Rev. A}\ }\textbf {\bibinfo {volume} {107}},\ \bibinfo
  {pages} {053320} (\bibinfo {year} {2023})}\BibitemShut {NoStop}%
\bibitem [{\citenamefont {Deffner}\ and\ \citenamefont
  {Campbell}(2017)}]{Deffner_2017}%
  \BibitemOpen
  \bibfield  {author} {\bibinfo {author} {\bibfnamefont {S.}~\bibnamefont
  {Deffner}}\ and\ \bibinfo {author} {\bibfnamefont {S.}~\bibnamefont
  {Campbell}},\ }\href {https://doi.org/10.1088/1751-8121/aa86c6} {\bibfield
  {journal} {\bibinfo  {journal} {Journal of Physics A: Mathematical and
  Theoretical}\ }\textbf {\bibinfo {volume} {50}},\ \bibinfo {pages} {453001}
  (\bibinfo {year} {2017})}\BibitemShut {NoStop}%
\end{thebibliography}
\end{document}